\documentclass[conference, a4paper]{IEEEconf}
\input epsf

\usepackage{graphicx}
\usepackage{multirow}
\usepackage{titlesec}
\usepackage{balance}
\usepackage{stfloats}
\usepackage{xcolor}
\usepackage{enumitem}
\usepackage{longtable}
\usepackage{booktabs}
\usepackage{csquotes}
\usepackage[compatibility=false]{caption}
\usepackage{subcaption}
\usepackage{tikz}       
\usetikzlibrary{external}
\usetikzlibrary{patterns}
\usepackage[framemethod=TikZ]{mdframed}
\usepackage{pgfplots}
\pgfplotsset{compat=1.18}
\usepgfplotslibrary{statistics}
\usepackage{array}

\usepackage{fontspec}

\usepackage[numbers,sort&compress]{natbib}
\usepackage{ragged2e,etoolbox}
\usepackage{orcidlink}

\usepackage{fontspec}

\newfontfamily\mheart{MHeart8.otf}[Path=fonts/]

\definecolor{myblue}{RGB}{180,208,244}
\definecolor{mycasebg}{RGB}{100, 80, 150}
\definecolor{OUred}{rgb}{0.855,0.031,0.071}
\definecolor{TILEBlue}{RGB}{0,52,102}

\definecolor{OUgreen}{rgb}{0.0,0.45,0.35} 
\definecolor{OUlightgreen}{rgb}{0.7,0.85,0.8} 

\renewcommand\thesection{\arabic{section}}
\renewcommand\thesubsectiondis{\thesection.\arabic{subsection}}
\renewcommand\thesubsubsectiondis{\thesubsectiondis.\arabic{subsubsection}}

\usepackage[acronym,nogroupskip,nonumberlist]{glossaries}

\newacronym{4C/ID}{4C/ID}{Four-component instructional design}
\newacronym{ACM}{ACM}{Association for Computing Machinery}
\newacronym{AI}{AI}{Artificial Intelligence}
\newacronym{API}{API}{Application Programming Interface}
\newacronym{APC}{APC}{Article Publishing Charge}
\newacronym{BoK}{BoK}{Body of Knowledge}
\newacronym{CER}{CER}{Computing Education Research}
\newacronym{CD}{CD}{Continuous Delivery}
\newacronym{CI}{CI}{Continuous Integration}
\newacronym{CI/CD}{CI/CD}{Continuous Integration and Continuous Delivery}
\newacronym{COVID-19}{COVID-19}{COrona VIrus Disease 2019}
\newacronym{CPNITS}{CPNITS}{Creative Programming for Non-IT-Students}
\newacronym{CRAAP}{CRAAP}{Currency, Relevance, Authority, Accuracy, Purpose criteria}
\newacronym{CS}{CS}{Computer Science}
\newacronym{CSEd}{CSEd}{Computer Science Education research}
\newacronym{CSEET}{CSEE\&T}{Conference on Software Engineering Education and Training}
\newacronym{CTL}{CTL}{Critical Testing Literacy}
\newacronym{DBE}{DBE}{Design Based Education}
\newacronym{DBLP}{DBLP}{DataBase systems and Logic Programming}
\newacronym{DBR}{DBR}{Design-Based Research}
\newacronym{DBER}{DBER}{Discipline-Based Education Research}
\newacronym{EACEA}{EACEA}{Euro\-pean Union Education and Culture Executive Agency}
\newacronym{EASE}{EASE}{The International Conference on Evaluation and Assessment in Software Engineering}
\newacronym{ECTS}{ECTS}{European Credit Transfer and Accumulation System}
\newacronym{ENACTEST}{ENACTEST}{European Innovation Alliance for Testing Education}
\newacronym{ES3C}{ES3C}{European Summer School on Science Communication}
\newacronym{FTE}{FTE}{Full Time Equivalent}
\newacronym{GAMEX}{GAMEX}{Game Experience Questionnaire}
\newacronym{GBL}{GBL}{Game-Based Learning}
\newacronym{GDPR}{GDPR}{General Data Protection Regulation}
\newacronym{GenAI}{GenAI}{Generative Artificial Intelligence}
\newacronym{GIPGUT}{GIPGUT}{Gamification Intellij Plugin for GUi Testing}
\newacronym{GUI}{GUI}{Graphical User Interface}
\newacronym{HBO-i}{HBO-i}{Foundation for Informatics Programmes of Dutch Universities of Applied Sciences}
\newacronym{HCI}{HCI}{Human-Computer Interaction}
\newacronym{ICAB}{ICAB}{InnovatieCentra Academisch Betàonderwijs}
\newacronym{ICMJE}{ICMJE}{International Committee of Medical Journal Editors}
\newacronym{ICSE}{ICSE}{International Conference on Software Engineering}
\newacronym{ICST}{ICST}{IEEE International Conference on Software Testing, Verification, and Validation}
\newacronym{IDE}{IDE}{Integrated Development Environment}
\newacronym{IEEE}{IEEE}{Institute of Electrical and Electronics Engineers}
\newacronym{IHEC}{IHEC}{Innovating Higher Education Conference}
\newacronym{IMC}{IMC}{Informatics Mortarboard Conference}
\newacronym{IPA}{IPA}{Institute for Programming Research and Algorithmics}
\newacronym{IST}{IST}{Information and Software Technology}
\newacronym{ISTQB}{ISTQB}{International Software Testing Qualifications Board}
\newacronym{ITiCSE}{ITiCSE}{Innovation and Technology in Computer Science Education}
\newacronym{JNDI}{JNDI}{Java Naming and Directory Interface}
\newacronym{JSON}{JSON}{JavaScript Object Notation}
\newacronym{K12}{K12}{Kindergarten through 12th grade (ages \~5-18)}
\newacronym{LEARNER}{LEARNER}{evaLuation assEssment softwARe eNgineering Education tRaining}
\newacronym{LM-GM}{LM-GM}{Learning Mechanics-Game Mechanics}
\newacronym{MC/DC}{MC/DC}{Modified Condition/Decision Coverage}
\newacronym{MOOC}{MOOC}{Massive Open Online Course}
\newacronym{MoT}{MoT}{Ministry of Testing}
\newacronym{NIOC}{NIOC}{Nederlands Informatica Onderwijs Congres}
\newacronym{NRO}{NRO}{The Netherlands Initiative for Education Research}
\newacronym{NWO}{NWO}{Dutch Research Council}
\newacronym{OER}{OER}{Open Educational Resources}
\newacronym{ORCID}{ORCID}{Open Researcher and Contributor ID}
\newacronym{OU}{OU}{Open Universiteit}
\newacronym{OUrsi}{OUrsi}{Open Universiteit Research Seminar Informatica}
\newacronym{OWASP}{OWASP}{Open Web Application Security Project}
\newacronym{PBL}{PBL}{Problem-Based Learning}
\newacronym{PRISMA}{PRISMA}{Preferred Reporting Items for Systematic Reviews and Meta-Analyses}
\newacronym{QPED}{QPED}{Quality-focused Programming Education}
\newacronym{RDW}{RDW}{The Netherlands Vehicle Authority}
\newacronym{RIMGEN}{RIMGEN}{Replicate, Isolate, Maximize, Generalize, and Externalize}
\newacronym{RST}{RST}{Rapid Software Testing~\copyright}
\newacronym{RTC}{RTC}{Romanian Testing Conference}
\newacronym{RUN-EU}{RUN-EU}{Regional University Network — European University}
\newacronym{SCQA}{SCQA}{Situation, Complication, Question, Answer}
\newacronym{SDT}{SDT}{Self-Determination Theory}
\newacronym{SE}{SE}{Software Engineering}
\newacronym{SEN}{SEN}{National Symposium Software Engineering The Netherlands}
\newacronym{SEET}{SEET}{Software Engineering Education and Training}
\newacronym{SIG}{SIG}{Special Interest Group}
\newacronym{SIGCSE}{SIGCSE}{Special Interest Group Computer Science Education}
\newacronym{SIGCSE-TS}{SIGCSE-TS}{Special Interest Group Computer Science Education Technical Symposium}
\newacronym{SLR}{SLR}{Systematic Literature Review}
\newacronym{SOLO}{SOLO}{Structure of observed learning outcome}
\newacronym{STGT}{STGT}{Socio-Technical Grounded Theory for Software Engineering}
\newacronym{SUS}{SUS}{System Usability Scale}
\newacronym{SUT}{SUT}{System Under Test}
\newacronym{SWEBOK}{SWEBOK}{Software Engineering Body of Knowledge}
\newacronym{TAM}{TAM}{Technology Acceptance Model}
\newacronym{TCM}{TCM}{TestCompass Model}
\newacronym{TDD}{TDD}{Test Driven Development}
\newacronym{TDL}{TDL}{Test Driven Learning}
\newacronym{TILE}{TILE}{Test Informed Learning with Examples}
\newacronym{UMUX}{UMUX}{Usability Metric for User Experience}
\newacronym{UPV}{UPV}{Universitat Politècnica de València}
\newacronym{VERSEN}{VERSEN}{Dutch National Association for Software Engineering}
\newacronym{VOR}{VOR}{Netherlands Educational Research Association}
\newacronym{WOZ}{WOz}{Wizard-of-Oz}
\glsdisablehyper{}

\AtBeginDocument{\setlength{\parindent}{1.5em}}

\newcommand{\papertitle}{Learning Critical Testing Literacy Through Puzzles:\\ an Experience Report}

\usepackage[]{hyperref}
\hypersetup{
    pdftitle={Learning Critical Testing Literacy Through Puzzles: an Experience Report},
    pdfauthor={Doorn, Niels; Knaack, Bart Th.; Vos, Tanja E.J.; Marín, Beatriz},
    pdflang={en-GB},
    pdfkeywords={education; software testing; didactic approach; puzzle-based learning; critical tester},
    colorlinks=false,
    linkbordercolor=white,
    urlbordercolor=white,
    pdfborder={0 0 0}
}

\usepackage{amssymb}
\usepackage{mdframed}

\definecolor{myblue}{RGB}{180,208,244}
\definecolor{mycasebg}{RGB}{100, 80, 150}
\definecolor{OUred}{rgb}{0.855,0.031,0.071}
\definecolor{TILEBlue}{RGB}{0,52,102}
\definecolor{P4TestBlue}{RGB}{147,201,231}
\definecolor{P4TestPink}{RGB}{242,196,213}

\definecolor{OUgreen}{rgb}{0.0,0.45,0.35} 
\definecolor{OUlightgreen}{rgb}{0.7,0.85,0.8} 

\newenvironment{takeaway}%
  {\begin{mdframed}[roundcorner=10pt,skipabove=8pt,skipbelow=0pt,backgroundcolor=P4TestBlue!55,nobreak=true]%
   $\blacktriangleright$\enspace\textbf{Takeaway: }}%
  {\end{mdframed}}

\begin{document}

\title{\papertitle}

\author{
    \IEEEauthorblockN{
        Niels Doorn$^{\dagger}$~\orcidlink{0000-0002-0680-4443} \quad
        Bart Th. Knaack~\orcidlink{0009-0001-2907-0614}
    }
    \IEEEauthorblockA{
        \textit{Open Universiteit},
        Heerlen, The Netherlands\\
        $^{\dagger}$Also: \textit{Iselinge Hogeschool}, Doetinchem, The Netherlands
    }
    \and
    \IEEEauthorblockN{
        Tanja E.J. Vos$^{\ddagger}$~\orcidlink{0000-0002-6003-9113} \quad
        Beatriz Marín~\orcidlink{0000-0001-8025-0023}
    }
    \IEEEauthorblockA{
        \textit{Universitat Politècnica de València},
        Valencia, Spain\\
        $^{\ddagger}$Also: \textit{Open Universiteit}, Heerlen, The Netherlands
    }
}

\maketitle

\begin{abstract}
In this paper, we report our experiences and takeaways from workshops using puzzles to learn \gls{CTL}. 

\noindent\textbf{Background:} Software testing is important, yet difficult to teach. In previous work, we introduced a \gls{BoK} of puzzle-based learning activities designed to teach \gls{CTL},
based on a model of critical tester's cognition.
This led to the development of a pedagogical framework called P4TEST for teaching software testing.
We used the puzzles and the framework to conduct thirteen workshops with students, testers, teachers, and primary school pupils to assess the validity of using puzzles as part of teaching critical testing literacy.

\noindent\textbf{Experience:} 
During the first eleven workshops, we used a semi-structured approach to conduct these workshops, but varied the puzzles, the use of materials, and the amount of time for the various parts of the workshop. 
After eleven workshops, besides mere observation, we introduced workbooks and think-aloud sessions into two more workshops to obtain more data about the learning experience.

\noindent\textbf{Observations:} Participants consistently perceived themselves as experimenting while solving the puzzles.
Students and professionals approached the same puzzles differently: students tended to converge on a solution, while professionals more often continued exploring the solution space.
Emotions were visible in behaviour but difficult to surface through written reflection alone.
Think-aloud sessions surfaced immediate reasoning; written reflections elicited more meta-cognitive reflection.
Directly linked to P4TEST, the theme \textit{Sensemaking / reflection-in-action}, captured how participants framed problems, navigated dead ends, and shifted strategies.

\noindent\textbf{Reflections:} Puzzles are not the intervention: the entire sequence of solving, debriefing, and reflecting is.
Designing that sequence more deliberately is the work ahead.
To reduce the friction of paper workbooks, we also developed an open-source web application with built-in analytics and configuration options to customise the workshops.
\end{abstract}
\IEEEoverridecommandlockouts
\vspace{1.5ex}

\IEEEpeerreviewmaketitle

\section{Introduction}
\label{puzzle:sec:introduction}

Software testing is an important but difficult skill to teach, and many educational approaches struggle to support it effectively.

Previous work has shown that students particularly struggle with the sensemaking processes involved in testing~\cite{DOORN2023102199}.

In response, earlier research introduced a Body of Knowledge (BoK) of puzzle-based learning activities to develop what they recognise as the five fundamental skills for sensemaking needed to become an effective software tester: experimentation, modelling, thinking, communicating and feeling~\cite{doorn2025puzzle}.

Building on this foundation, a pedagogical framework (P4TEST) was developed to structure the teaching of software testing as a process of sensemaking that includes the development of models,
exploration, and reflection-in-action~\cite{vos2025P4TEST}.

Although these works~\cite{doorn2025puzzle, vos2025P4TEST} define relevant competences and propose instructional structures for software testing, and prior research on puzzle-based learning shows that puzzles can be effective for learning \cite{FSM12}, there is still limited understanding of how puzzle-based learning specifically supports the development of software testing competences in practice.

This paper explores
this gap by reporting on the experiences of running puzzle-based workshops and examining how participants engage with these activities.

The puzzles used come from the mentioned \gls{BoK}~\cite{doorn2025puzzle} and are designed to evoke testing-related sensemaking behaviour and critical, lateral, horizontal and vertical thinking.
The puzzles are not programming exercises: they are domain-agnostic problems --- such as number sequences with multiple valid continuations, spatial reasoning tasks with implicit constraints, and logic puzzles with seemingly insufficient information --- chosen because they demand similar cognitive moves as software testing: forming hypotheses, questioning assumptions, exploring systematically, and deciding when evidence is sufficient.

Each puzzle has more than one
solution.
This design compels students not to settle for the first answer, but to ask clarifying, assumption-challenging, evidence-seeking, and counterfactual questions, and to explore alternative solutions.
The puzzles try to evoke multiple `aha!' moments of insight~\cite{sternberg_nature_1995} that create engagement and deepen understanding, making learning both more meaningful and memorable.
The puzzles are easy to remember to increase learning retention.

The development of these puzzles as an educational intervention was done iteratively and incrementally.
What started as a small idea led to a more solid approach through workshops with various types of participants, including professionals, teachers, students, and elementary school pupils.

\begin{figure*}[ht!]
   \centering
   \includegraphics[width=0.8\linewidth]{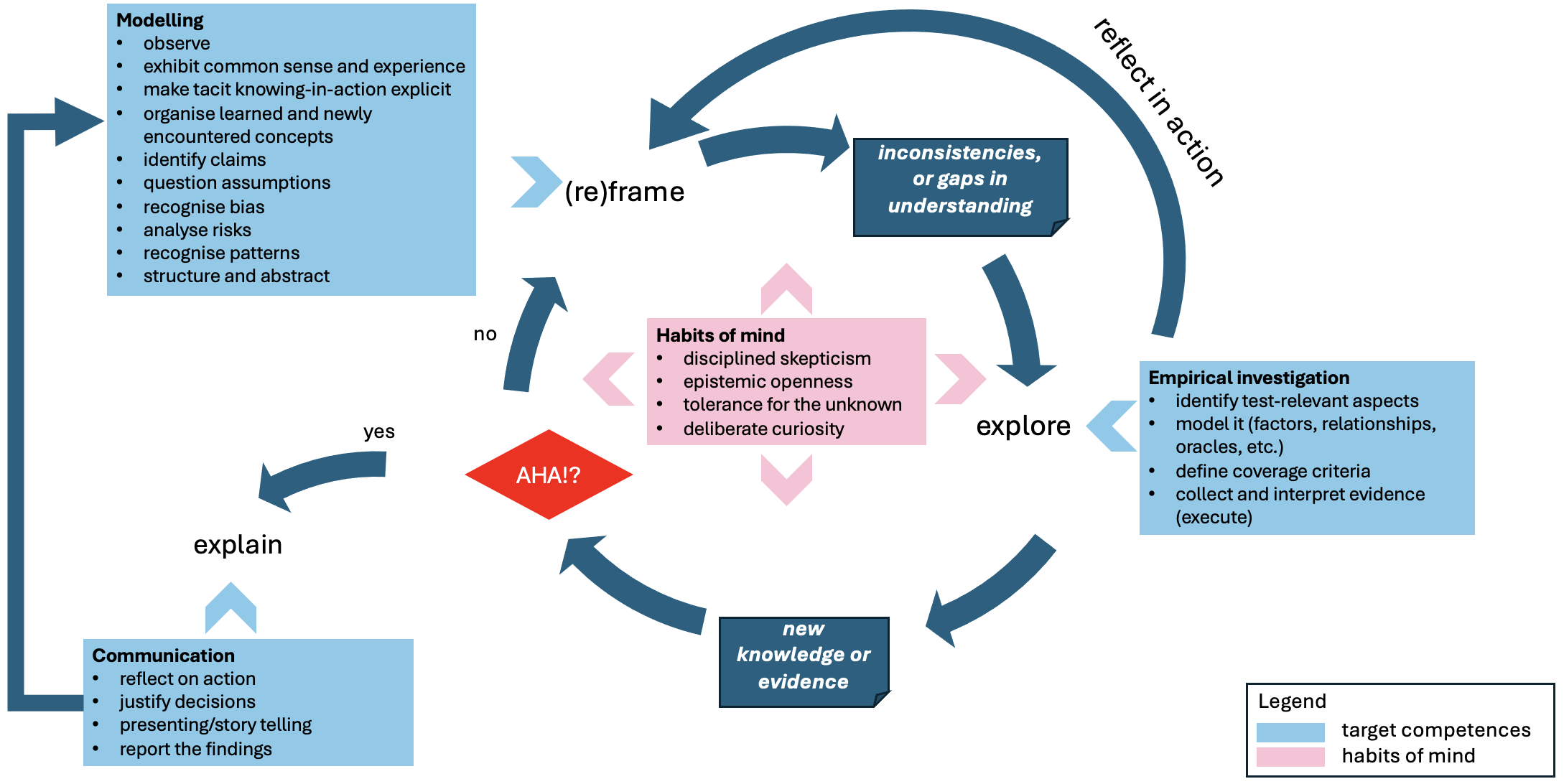}
   \caption{P4TEST Pedagogical framework to teach software testing~\cite{vos2025P4TEST}}
   \label{fig:P4TEST}
\end{figure*}

We designed the workshops using the P4TEST framework as shown in Figure~\ref{fig:P4TEST} to observe what happens when participants try to solve these puzzles in an educational and training settings.

We ran the workshops in different contexts
and we learned from each workshop.

At some point, our observations raised questions that we could not answer by observation from the front of the room alone:

\begin{itemize}
    \item are students and professionals experiencing the puzzles the way we intend
    \item is what we are seeing actually aligned with the aspects of the P4TEST framework that motivated the design of the puzzles and workshops.
\end{itemize}

To explore these questions, we looked at ways to gain more insight and decided to start collecting written reflections and recordings of think-aloud sessions.
We integrated the written reflections using workbooks into the workshops and conducted separate think-aloud sessions.
The workbooks were structured according to P4TEST themes with the addition of experiences and emotions themes.
Each puzzle had the same Likert scale and open questions.

In this article, we describe what we learned from the workshops.
The questions those experiences raised and what we did to investigate them come next.
This is followed by a description of what we found and what the implications can be for future workshops.
As a practical contribution, we released an open-source web application~\cite{doorn_2026_19426383} that provides a digital alternative to the paper workbooks we used in the last two workshops, with built-in analytics and export functionality for researchers who run similar workshops.

\section{The puzzles we used in the workshops}
Six puzzles were selected from the \gls{BoK} of puzzle-based learning activities for software testing.

\textbf{The ``Next-line'' puzzle.}
Participants are shown the beginning of a number sequence like the following:

\begin{minipage}{\linewidth}
\vspace{0.5em}
{\small
\begin{verbatim}
1
1 1
2 1
1 2 1
1 2 1 1
\end{verbatim}
}
\vspace{0.5em}
\end{minipage}

The most common interpretation of this sequence is the \emph{look-and-say} sequence, where each line describes the preceding one (e.g. describing what is on line two ``two ones'' leads to line three becoming \texttt{2\ 1}).
However, the puzzle has multiple possible answers: students have proposed rules based on the Fibonacci sequence, Morse code, and polynomial functions.
The puzzle develops pattern recognition, hypothesis formulation, and tolerance for ambiguity.
Participants must decide when a proposed rule is sufficiently supported and remain open to alternative explanations even after reaching a plausible answer.

\textbf{The ``Nine-dots'' puzzle.}
Participants must connect all nine dots on a 3$\times$3 grid using as few straight line segments as possible, without lifting the pen~\cite{Loyd1914}.
The puzzle is a well-known illustration of how implicit self-imposed constraints block problem-solving~\cite{Adams2001}, and is often cited as the origin of the phrase ``thinking outside the box''~\cite{deBono1970}.
It is important to note that they are dots, not points, and lines do not need to go through the centre.
The learning goals of this puzzle are to create awareness of hidden assumptions, support lateral thinking, and demonstrate the value of persistence when an approach fails.

\textbf{The ``ABC-connect'' puzzle.}
Participants must connect three pairs of matching letters (A-A, B-B, C-C) on a grid without crossing any lines.
It develops the ability to switch approaches when a first attempt fails, challenges first impressions, and rewards systematic exploration.

\textbf{The ``Three sons'' puzzle.}
Also known as the Census-Taker Problem~\cite{MS90}, this logic puzzle asks participants to determine the ages of three children using seemingly insufficient information: their ages total to~13, their product equals the age of person~B, and the oldest child weighs 30~kg.
Person~B's statement that the first two clues are insufficient is itself a crucial clue.
Learning goals include tolerance for the unknown, systematic enumeration of possibilities, construction of meaning from unrelated clues, and questioning assumptions.

\textbf{The ``Weird symbols'' puzzle.}
Participants are shown a sequence of mirror-symmetrical symbols such as: {\mheart K 1 L 2 M 3}\footnote{{\mheart K 1 L 2 M 3} is mirror-symmetrical K 1 L 2 M 3, the next symbols could be {\mheart N 4} (counting) or {\mheart N 5} (primes)\ldots}, and asked to identify the next symbol~\cite{Meyeretal2014}.
The puzzle has multiple valid solutions (arithmetic, geometric, phonetic, etc.).

Its learning goals focus on systematic pattern recognition using multiple heuristics and on equivalence partitioning: grouping symbols by shared properties (e.g.\ symmetry, number of curves) to find structure.

\textbf{The ``Dice puzzle''.}
Participants get a set of distinct dice, for example, two six sided, a twenty sided, and a dice with colours.
The participants throw the dice and ask the facilitator what the outcome of their roll is.
The facilitator uses a secret algorithm to determine the outcome that could take some of the dice into account, ignore others, and include other non-dice related variables such as the current time, or the number of hands on the table.
The participants need to figure out what the algorithm is.
Learning goals are to challenge assumptions and using a structured approach.

\section{Workshop experiences}
\label{puzzle:sec:experiences}

In this section, we describe our anecdotal experiences from the workshops, that were given.
All workshops started with an introduction to give participants an understanding of the format of the workshop and the goal of the puzzles.
This was followed by a practice round with the first puzzle, which was solved and discussed plenary.
Following the introductory puzzles, one by one new puzzles were presented and the participants were given time to come up with solutions.
Participants were facilitated with materials such as pen and paper, print outs, glue, and scissors.
Each puzzle was followed by a debrief in which possible solutions found by the participants were discussed and other solutions were presented.
During the debrief, we also reflected on the assumptions made, the influence of biases, and the previous experiences of the participants, and linked the puzzles to software testing practices.
\\
\\
\noindent \textbf{Workshop 1}

\noindent \textbf{where}: \gls{OU} study day, Eindhoven

\noindent \textbf{when}: 7$^{th}$ of December 2024

\noindent \textbf{what}: an extracurricular activity that students can choose from a range of parallel events

\noindent \textbf{audience}: students enroled in Bachelor’s or Master's programmes in Computer Science

\noindent \textbf{experience}: Since the \gls{OU} offers part-time
learning, most of these students also work
.

During this workshop, we tried the Nine-dots, Three sons, and the Dice puzzle.
The participants clearly enjoyed them and were reluctant to leave once the allocated session time had ended.
They rated the session highly afterwards.

\begin{takeaway}
Participants reported enjoying learning through the puzzles, suggesting that this approach is a promising candidate for further investigation as a learning activity.
\end{takeaway}

The introduction to this workshop was relatively long, trying to explain the connection between the puzzles and software testing.
This left us less time to work on the puzzles and debriefs.

However, our observations indicate that the debrief played a key role in the learning process, as it prompted participants to reflect on and articulate their reasoning.
This aligns with Kolb's experiential learning cycle~\cite{Kolb1984}, where reflection is essential to consolidate experience into learning.

\begin{takeaway}
Keep the introduction short and dive directly into the puzzles. Take enough time for the debriefs, since that is where learning and \textit{aha!} moments happen.
\end{takeaway}

\noindent \textbf{Workshop 2}

\noindent \textbf{where}: NHL Stenden University of Applied Sciences

\noindent \textbf{when}: 11$^{th}$ of February 2025

\noindent \textbf{what}: part of a first year testing course from the Bachelor of Information Technology programme

\noindent \textbf{audience}: bachelor students

\noindent \textbf{experience}: The workshop was divided into two groups, each attending a separate session in sequence.
The first group generated solutions quickly, in some cases proposing novel approaches.
However, once a solution was found, they tended to fixate on it and move on rapidly to the next puzzle.

In contrast, the participants in the second session demonstrated greater contemplation and reflection on their solutions.
Despite both sessions following the same format and involving a similar audience, the participants engaged with the puzzles in noticeably different ways.

\begin{takeaway}
Even with a similar audience and identical format, group dynamics can significantly influence how participants approach and engage with puzzles. Facilitators should therefore be prepared to adapt to these differences during the session.
\end{takeaway}

\noindent \textbf{Workshop 3}

\noindent \textbf{where}: \gls{ICAB}, Maastricht

\noindent \textbf{when}: 13$^{th}$ of March 2025

\noindent \textbf{what}: a conference for educators

\noindent \textbf{audience}: teachers in various programmes of different universities

\noindent \textbf{experience}:
Unlike previous workshops, where pen and paper were readily available and commonly used, participants in this session at first did not make a practical start solving the puzzles on paper. Instead, many attempted to solve the puzzles mentally, which led to several participants becoming stuck early in the process.

This contrast with earlier sessions suggested that the availability and use of pen and paper played a more important role than we had initially assumed.
In response, we began to explicitly encourage participants in subsequent workshops to externalise their thinking by writing, sketching, or experimenting on paper.

\begin{takeaway}
Providing and explicitly encouraging the use of pen and paper helps participants externalise their thinking, overcome fixation, and make progress in solving the puzzles.
\end{takeaway}

\noindent \textbf{Workshop 4}

\noindent \textbf{where}: EuroStar 2025, Edinburgh

\noindent \textbf{when}: 4$^{th}$ of June 2025

\noindent \textbf{what}: a conference for test professionals

\noindent \textbf{audience}: test professionals from all over the world

\noindent \textbf{experience}: The seating arrangement in the workshop room was cabaret-style, without tables for writing.
This made it difficult for participants to collaborate effectively in groups. As a result, the session took on a “popcorn” dynamic, where attendees called out ideas in response to a puzzle, but the discussion lacked structure and depth.
Additionally, the presence of a main stage created distance between the facilitators and the audience, making one-on-one interaction challenging.

\begin{takeaway}
Proper set up of the room is essential for running the workshop effectively.
\end{takeaway}

\noindent \textbf{Workshop 5}

\noindent \textbf{where}: SpaceRockIT festival, Huissen

\noindent \textbf{when}: 5$^{th}$ of September 2025

\noindent \textbf{what}: a conference for IT professionals

\noindent \textbf{audience}: IT professionals from The Netherlands

\noindent \textbf{experience}: As the name suggests, this was an outside conference in festival style.

We took an approach in which we did not try to do as many puzzles as the time slot allowed, but focused on the debrief in which we tried to get as many of the experiences, emotions, and epistemics from the participants as possible.
By sharing experiences in the group, we observed that this was valuable for all participants.
We also noticed in both this and previous sessions that the puzzles and the order in which they are presented influence each other.
Mainly the geospatial puzzles tend to influence each other; the Nine-dots seem to prime the participants for the ABC-connect and vice versa, showing more confident participants at the start of the second puzzle.
This could be related to the concept of conceptual priming~\cite{Sweller1976}: the first puzzle activates the same cognitive frame used to solve the second.
Blocking sequencing, the same category of puzzles completed one after the other, is known to result in lower performance compared to solving puzzles in an interleaved sequence~\cite{10.1007/978-3-642-30950-2_24}.

\begin{takeaway}
Puzzles within the same category could have the effect of conceptual priming.
Moreover, interleaved sequencing could result in an even better learning effect.
\end{takeaway}

\noindent \textbf{Workshop 6}

\noindent \textbf{where}: \gls{HBO-i} JobEvent

\noindent \textbf{when}: 2$^{nd}$ of October 2025

\noindent \textbf{what}: a yearly returning event for both students and teachers of the Dutch higher education \gls{CS} programmes

\noindent \textbf{audience}: Lecturers of \gls{CS} programmes

\noindent \textbf{experience}: The lecturers gave positive feedback on the puzzles and a lively discussion followed after trying the puzzles.
This session also further strengthened the idea that educators struggle with teaching software testing to their students.

\vspace{1em}
\noindent \textbf{Workshop 7}

\noindent \textbf{where}: Noordertest conference

\noindent \textbf{when}: 16$^{th}$ of October 2025

\noindent \textbf{what}: conference for test professionals

\noindent \textbf{audience}: test professionals from the North of the Netherlands

\noindent \textbf{experience}:

We observed behaviour similar to that in other sessions with test professionals. Some participants approached the puzzles with a healthy degree of scepticism, explicitly questioning how they might be misled by the facilitators.

As in previous sessions, the debrief generated the most valuable insights, with participants sharing diverse solution strategies and reasoning processes.

When comparing sessions with test professionals to those with students and teachers, professionals appeared to develop an understanding of the puzzles more quickly.

This may indicate that they already possess, to some extent, the cognitive skills associated with critical testing, such as curiosity, scepticism, and empirical investigation.

\begin{takeaway}
Test professionals appear to already exhibit several of the cognitive behaviours associated with critical testing. Doing puzzles may reinforce and make existing practices explicit.
\end{takeaway}

\noindent \textbf{Workshop 8}

\noindent \textbf{where}: Sint Willibrordusschool, Herveld, The Netherlands

\noindent \textbf{when}: 12$^{th}$ of November 2025

\noindent \textbf{what}: workshop at primary school

\noindent \textbf{audience}: primary school pupils

\begin{figure}
    \centering
    \begin{subfigure}{0.45\linewidth}
        \centering
        \includegraphics[width=\linewidth, trim=0 0 0 0, clip]{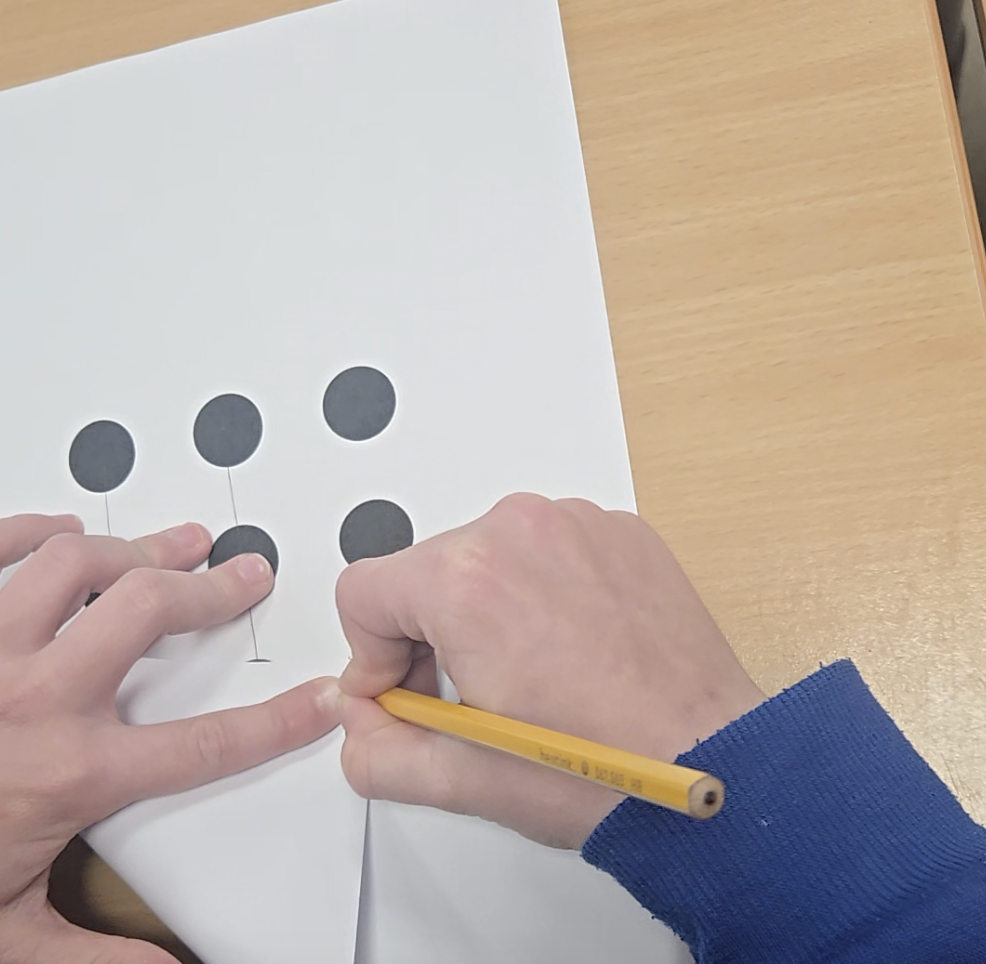}
        \caption{Using the paper to move the pen}
        \label{fig:willibrordA}
    \end{subfigure}
    \hfill
    \begin{subfigure}{0.45\linewidth}
        \centering
        \includegraphics[width=\linewidth, trim=0 0 0 20pt, clip]{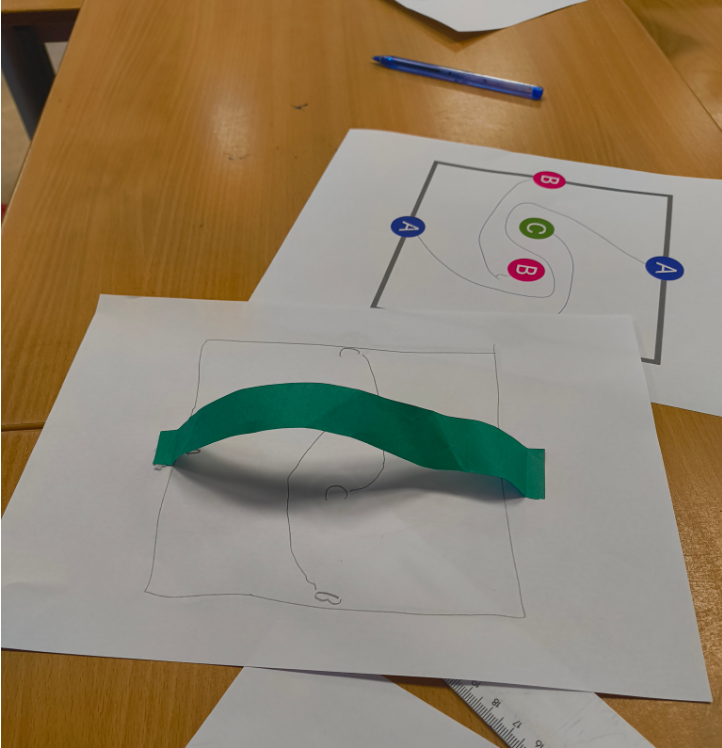}
        \caption{Glue and paper to make a connection}
        \label{fig:willibrordB}
    \end{subfigure}
    \caption{New solutions found by the primary school children}
    \label{fig:willibrord}
\end{figure}

\noindent \textbf{experience}: This particular school has a strong focus on an evidence-based way of working and teaching methods that stimulate learning through inquiry and experimentation.
The puzzles fit well into their programme.
We changed the tone of the way of presenting the puzzles slightly to avoid difficult terms or to include explanations of such terminology, but the puzzles remained unchanged.
The debrief was also adjusted to relate to the age of the pupils, but had a similar effect as in the other session by sharing ideas, emotions, and habits of mind with each other.
Some of the pupils took approaches we have not seen by other audiences, such as literally taking a different perspective by climbing on the table to look at a puzzle or by using glue and strips of paper to make connections on the puzzles.
In Figure \ref{fig:willibrord} we see two new approaches to the spatial puzzles. We see how the Nine-dots puzzle was solved by lifting and folding the corner of the paper to connect the dots without ever lifting the pen.

\begin{takeaway}
Although not originally designed for primary education, the puzzles appear to translate well to younger audiences, suggesting potential for adapting puzzle-based learning activities to support the development of inquiry and critical thinking skills at this level.
\end{takeaway}

We scheduled the Three sons puzzle at the end of the session.
The participants found this puzzle more difficult, leading to signs of mild frustration.
This response may be explained in part by the timing of the activity, as the participants had already been engaged in puzzle solving for some time.
However, it is possible that the puzzle itself, characterised by incomplete and ambiguous information, brought the tolerance of participants to the unknown to the surface.
This created an opportunity to recognise these reactions and reflect on how to deal with them.

In later workshops, we used this more explicitly as a moment to discuss uncertainty and how participants respond to it. \footnote{maybe this puzzle is just not fit for younger audiences since it is already challenging for adult audiences and is less playful than the other puzzles.}

\begin{takeaway}
Puzzles can make participants' responses to uncertainty visible, providing opportunities to reflect on and develop tolerance for the unknown.
\end{takeaway}

\noindent \textbf{Workshop 9}

\noindent \textbf{where}: Agile Testing days 2025

\noindent \textbf{when}: 27$^{th}$ of November 2025

\noindent \textbf{what}: a conference for test professionals

\noindent \textbf{audience}: test professionals from all over the world

\noindent \textbf{experience}: The room was organised into eight tables of up to six participants each, providing ample opportunity for communication and collaboration.
This setup led to a form of informal competition between tables, with participants actively comparing approaches and attempting to outperform each other.

Similar to our observations at Noordertest, test professionals demonstrated the habits of mind from P4TEST.
This further supports our observation that test professionals already possess many of the habits of mind that the puzzles aim to develop.

\vspace{1em}
\noindent \textbf{Workshop 10}

\noindent \textbf{where}: Promis-ES event

\noindent \textbf{when}: 29$^{th}$ of November 2025

\noindent \textbf{what}: a yearly event for PhD students from the Science Faculty of the Open Universiteit

\noindent \textbf{audience}: PhD students and professors from the Science Faculty

\noindent \textbf{experience}:
The participants enjoyed the session, but we did not gain much new knowledge.
Still, to be complete, we decided to include this experience.

\vspace{1em}
\noindent \textbf{Workshop 11}

\noindent \textbf{where}: MoT Valencia

\noindent \textbf{when}: 18$^{th}$ of December 2025

\noindent \textbf{what}: Software Testing Meetup

\noindent \textbf{audience}: Professional Software Testers

\noindent \textbf{experience}:
During this session, one participant showed a strong discomfort with uncertainty when working on the Three sons puzzle.
This influenced group dynamics and made the challenge of dealing with incomplete or ambiguous information more visible.

This provided an opportunity to explicitly address how participants respond to uncertainty and how such reactions can be reframed as part of the learning process.
It also reinforced earlier observations that puzzle-based activities can help surface differences in tolerance for the unknown, which can be used as a starting point for reflection and discussion.

\section{A closer look at what happens during the puzzles}
\label{puzzle:sec:empirical}

The above takeaways emerged from direct observation during these eleven workshops and discussions during the debriefs.
They are anecdotal: we noticed things, talked about them with participants, and adjusted our practice.

But at some point, these observations raised questions we could not answer by merely watching from the front of the room.

We observed that students and professionals seemed to approach the puzzles differently, but how did they themselves describe what was happening?
We noticed behaviours that looked like hypothesis testing, pattern recognition, and frustration-driven strategy shifts, but could we actually see these in a more systematic way?
And how well did what we observed and what the participants reported align with the P4TEST framework?

These are the kinds of questions teachers ask when they want to understand whether their teaching is doing what they intended.
To explore them, we conducted two additional workshops with more deliberate data collection: one with a class of undergraduate students as part of an existing software testing course and one with a group of expert testers during a meetup event.
What we did differently in those workshops was small but deliberate: we asked participants to work through a printed workbook so that we could see not only their solutions, but also their reasoning, and we invited some pairs of students to work on puzzles while thinking aloud so that we could observe the process more directly.

\section{What we did}
\label{puzzle:sec:design}

In both workshops, we followed the same workshop structure described above: an introduction, puzzle solving, a plenary debrief, and we added time for reflection.
What we added, in order to look more closely at what was happening, was a paper workbook that each participant completed during the session and -- for the student session -- the option to take part in a separate think-aloud session afterwards.

\subsection{Audience}
Seven bachelor of computer science students of a University of Applied Sciences took part in the students' workshop.
Of these, four participated in a follow-up think-aloud session in two groups of two students.
The workshop with test professionals was attended by 13 participants, of whom 12 consented to participate in the study.
These professionals work in various organisations as software testers.
In this study, no data on their demographics and background was collected; however, it is known that these professionals are a heterogeneous group of people.
Participation was voluntary in all cases and did not affect students' grades in any way.

\subsection{How we collected data}

\subsubsection{Workbook sessions}
\label{puz:subsec:workbooks}

For each puzzle, the workbook included:

\begin{itemize}
    \item space for describing the solutions found;
    \item a structured reflection section covering the target competencies and habits of mind of P4TEST, i.e. knowledge curation/modelling, experimenting/empirical investigation, communication, and disposition \footnote{In the workbooks we used Disposition to denote the collective of Habits of Mind, to elevate the need to explain the Habits of Mind} (see Table~\ref{tab:likert-items});

    \item a reflection item on \textit{Experience}, included to account for differences in prior knowledge and background between participants (e.g., students versus professionals), enabling us to interpret variations in problem-solving behaviour in light of prior exposure;
    \item a reflection item on \textit{Emotions}, included to capture the affective aspect of puzzle solving, building on the notion of the critical tester as involving `Feelings'~\cite{doorn2025puzzle}.

    This aspect could enable us to examine how factors such as frustration, bias, confidence, and motivation influence reasoning processes and decision-making during the tasks.
\end{itemize}

Each aspect was operationalised through a short Likert-scale item (1 = strongly disagree, 5 = strongly agree) and an open question. The same six aspects were used for each puzzle (see Table~\ref{tab:likert-items}).
\begin{table}[!t]
\caption{Reflection Likert items used for each puzzle}
\label{tab:likert-items}
\centering
\small
\setlength{\extrarowheight}{-2pt}
\renewcommand{\arraystretch}{0.85}
\begin{tabular}{lp{4.5cm}}
\toprule
\textbf{Aspect} & \textbf{Statement} \\
\midrule
Experimenting      & I systematically tried different ideas and tested my assumptions. \\
Knowledge Curation & I organised clues, patterns, and partial results in a structured way. \\
Communicating      & I could clearly explain my reasoning and evidence to someone else. \\
Disposition        & I stayed curious and sceptical instead of jumping to conclusions. \\
Experience         & In this puzzle, I relied on my previous experience with similar tasks. \\
Emotions           & My feelings (e.g.\ frustration, excitement, confidence) influenced my decisions. \\
\bottomrule
\end{tabular}
\end{table}

Each puzzle followed the same procedure:
\begin{enumerate}[nosep]
  \item \textbf{Introduction (2 min).} The instructor presented the puzzle and clarified that multiple approaches were valid.
  \item \textbf{Puzzle solving (10--15 min).} Participants worked individually or in small groups without hints from the instructor.
  \item \textbf{The whole class debrief (5 min).} The group briefly discussed approaches, difficulties, and solution strategies.
  \item \textbf{Reflection (5 min).} Participants completed the Likert items and open-ended prompts for that puzzle.
\end{enumerate}

This gave us for each puzzle and for each participant a written solution, a Likert score, and an open-ended reflection for each aspect.

\subsubsection{Think-aloud sessions}
Think-aloud sessions were scheduled separately with two pairs of students.
Each pair was briefly introduced to the think-aloud method and completed a short warm-up task to get used to `thinking aloud' before working on two puzzles, the Three sons and the Weird symbols, while verbalising their thinking.
The researcher refrained from content-related help, only prompting ``keep talking'' if needed, and asked brief meta-questions at the end.
The sessions lasted around 20--30 minutes and were recorded in audio and video; written artefacts produced during the solving were also collected.

\subsection{How we analysed the data}
Likert items were analysed using basic descriptive statistics.
The purpose was to characterise how students self-report the relevance of different aspects across puzzles, rather than to perform inferential statistical testing.

The open-ended reflections and the think-aloud recordings and transcriptions were qualitatively analysed using a reflexive thematic analysis, following steps of familiarisation with the written responses, generating initial codes (both deductively from P4TEST and inductively from the data), searching for themes that capture recurring patterns in how students describe their experiences and strategies, and reviewing, defining and naming the topics related to our questions.

The competences and habits of P4TEST served as a guiding framework, but the coding remained open to emergent themes that may refine or extend the model.

\subsection{Ethical considerations}

Participation in the study was voluntary.
For the workbook sessions for students, the activities were embedded in normal teaching practice; students were informed that their anonymised responses might be used for research purposes and had the option to opt out without consequences for their coursework.

Written consent was required from both students and professionals.

The recordings and transcripts were anonymised; identification details were removed or masked.
Data were stored securely according to institutional and legal requirements (e.g. \gls{GDPR}), and access is restricted to the research team.

\subsection{Data availability}

All anonymised data that can be shared openly, including Likert scores, open-ended workbook reflections, and codebooks, are available in the research data set~\cite{doorn2026puzzledata}.

\section{What we found}
\label{puzzle:sec:results}

\subsection{Self-reported engagement (Likert-scale)}

Across the puzzles, participants consistently reported high levels of \textit{Experimenting}, with median scores of~4 for both students and professionals.
This indicates that the participants perceived themselves to actively try to evaluate different solutions during puzzle solving.

Students and professionals scored differently on several puzzles, particularly on \textit{Disposition}, \textit{Knowledge Curation}, and \textit{Communicating}.
For example, in the \textbf{Next-line} puzzle, students reported lower median scores than professionals on these aspects.
In contrast, the \textbf{Nine-dots} puzzle showed similar median scores across all aspects for both groups, suggesting comparable self-reported engagement for this puzzle.
For the \textbf{ABC-connect} puzzle, students reported relatively high scores across most aspects, while professionals reported lower scores on \textit{Emotions}.
These differences were not consistent between puzzles and are based on descriptive statistics, but these are varied across puzzles and should be explored further.

\textit{Emotions} tended to score lower for both groups (medians of~3 or less in most puzzles), indicating that participants did not strongly perceive their feelings as influencing their decisions or found it harder to articulate this aspect.

\begin{takeaway}
\textit{Emotions} are visible in behaviour but difficult to elicit through written reflection alone.
The debrief may be a better moment to surface them.
\end{takeaway}

\subsection{Themes in participant reflection}

The open-ended written reflections were coded using the competences and habits of mind from P4TEST, plus the
sensemaking process of the framework, i.e. framing, exploring, reflection-in-action.
Following the workbooks, two additional themes were used, \textit{Emotions} and \textit{Experience}.

Figure~\ref{fig:codeanalysis} shows the relative distribution of these themes in the five puzzles, separately for professionals and students, and for the two data collection methods.

For the \textit{Sensemaking} theme, we identified four codes: \textit{Framing the puzzle}, \textit{Dead ends}, \textit{Strategy shifts}, and \textit{Consolidating}.
Together, these capture how participants frame the puzzle, how they respond when progress comes to a halt, how they change their approach, and how they conclude working on a puzzle.

\subsubsection{Framing the puzzle}
\textit{Framing the puzzle} captures the moment the participants form an initial understanding of what kind of problem they are dealing with.
For the Next-line puzzle, one participant immediately reframed the task: ``It looks like a maths puzzle, but it is not'', opening up to non-mathematical solution paths before exploration had begun.
In the Weird symbols puzzle, a participant constructed an explicit working hypothesis: ``\ldots so whatever is gonna be, let's say it's gonna be symmetrical, yeah. Because there's a pattern, it's symbols, and it's a sequence.''
For the Three sons puzzle, one participant constrained the solution space: ``No one can be too old. So, let's say one person can be 11 and the other one's going to be one. Yeah, the other two will be one.''
These initial framings function as a scaffold for subsequent exploration: they narrow the search space before experimentation begins, but can also lead to exploring solution spaces that do not produce viable results.

\subsubsection{Dead ends}
\textit{Dead ends and fixation} capture moments where participants recognised that a path to a solution leads to nowhere, or that they had become stuck during their exploration.
In the Three sons puzzle, participants fixed on the mismatch between what the puzzle requires and what they had available: ``You need to know the age of Person B'' and ``I don't know the age of a child weighing 30 kilogrammes.''
A more subtle form of fixation appeared in the ABC-connect puzzle, where a participant wrote, ``I found a solution [\ldots] so no further experimentation''.
Reaching an acceptable answer became a form of closure that prohibited the search for further solutions.
The Nine-dots puzzle produced a similar pattern: ``I filled in the first thing that worked''.
In the Next-line puzzle, one participant reflected, ``I tried some ideas before, but I think I was getting further from the answer'', showing a realisation of moving into a direction that won't lead to a valid solution.

\subsubsection{Strategy shifts}

\textit{Strategy shifts} are deliberate redirections of an approach.
In the Nine-dots puzzle, one participant enumerated the alternatives they considered: ``Folding, rolling, drawing lines, drilling holes through the paper'', while another described the conceptual shift that gave better solutions: ``Drawing lines between points and re-organising were my main strategies. By connecting dots outside the box gave the option to use fewer lines.''
Another shift appeared in the ABC-connect puzzle, where a participant noted, ``Being less serious helps to think outside of the box'', suggesting that deliberately adjusting one's attitude can itself function as a strategy.

\subsubsection{Consolidating}

\textit{Consolidating} captures participants finishing their solution attempt, often leaving some doubt of the correctness or completeness of their found solutions.
A student after finding a possible solution to the Three sons puzzle reflected: ``And then this will make sense, I think''.

\subsection{Observations}

Across all puzzles and participant groups, \textit{Sensemaking / reflection-in-action} and \textit{Experimenting} are the most prominent themes, and \textit{Experimenting} was particularly high for the Weird symbols puzzle.
This puzzle seems to have the effect of alienating participants and forcing them into a trial-and-error approach.
\textit{Emotions} and \textit{Communicating} occupy the middle ground, while \textit{Knowledge Curation} and \textit{Disposition} are the least-coded categories throughout, indicating that participants rarely reflected on how they managed prior knowledge or on their general attitude toward puzzle-solving.

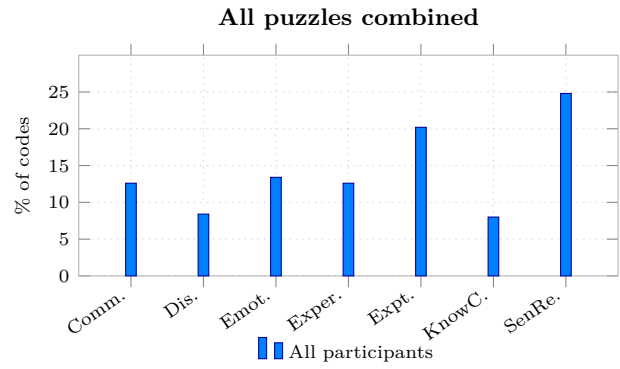
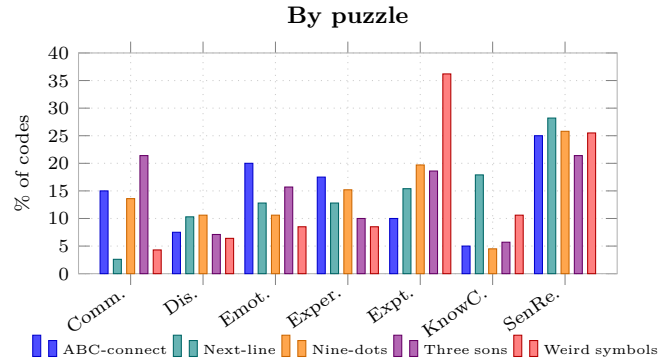
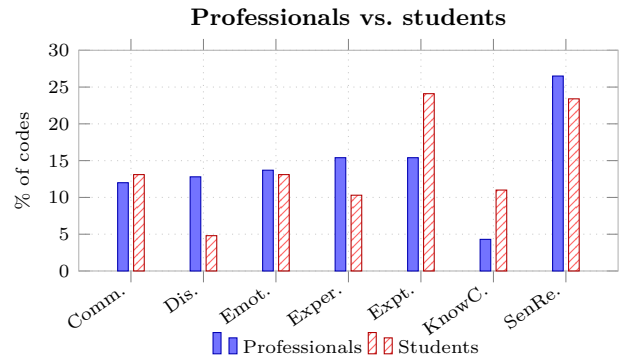
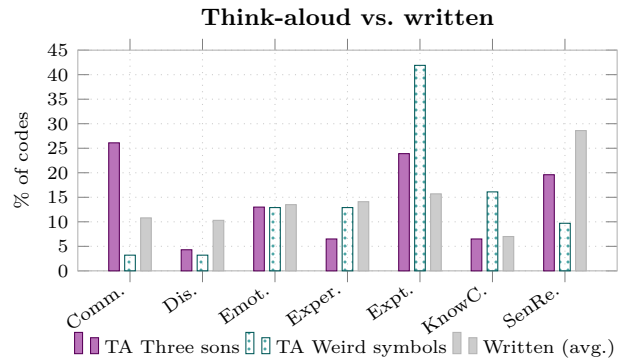
\begin{figure*}[ht!]
\centering
\pgfplotsset{
  codeaxis/.style={
    ybar,
    bar width=4pt,
    width=1\linewidth,
    height=4.5cm,
    enlarge x limits=0.12,
    symbolic x coords={Comm.,Dis.,Emot.,Exper.,Expt.,KnowC.,SenRe.},
    xtick=data,
    xticklabel style={font=\scriptsize, rotate=35, anchor=east},
    ymin=0, ymax=30,
    ytick={0,5,10,15,20,25},
    yticklabel style={font=\scriptsize},
    ylabel style={font=\scriptsize},
    ylabel={\% of codes},
    title style={font=\small\bfseries},
    legend style={font=\scriptsize, at={(0.5,-0.25)}, anchor=north,
                  draw=none, fill=none, legend columns=-1},
    grid=major,
    grid style={dotted, gray!40},
    axis line style={gray!60},
    clip=false,
  }
}

\begin{subfigure}[t]{0.48\linewidth}
\centering
\begin{tikzpicture}
\begin{axis}[
  codeaxis,
  title={All puzzles combined},
  legend entries={All participants},
]
\addplot[fill=blue!50!cyan, draw=blue!60!black]
  coordinates {
    (Comm., 12.6)
    (Dis.,  8.4)
    (Emot., 13.4)
    (Exper.,12.6)
    (Expt., 20.2)
    (KnowC., 8.0)
    (SenRe.,24.8)
  };
\end{axis}
\end{tikzpicture}
\caption{Relative code distribution across all five puzzles and all participants. \textit{Sensemaking / reflection-in-action} and \textit{Experimenting} dominate; \textit{Knowledge Curation} and \textit{Disposition} are least prominent.}
\label{fig:codes:all}
\end{subfigure}
\hfill

\begin{subfigure}[t]{0.48\linewidth}
\centering
\begin{tikzpicture}
\begin{axis}[
  codeaxis,
  title={By puzzle},
  legend entries={ABC-connect, Next-line, Nine-dots, Three sons, Weird symbols},
  legend style={font=\tiny, at={(0.5,-0.25)}, anchor=north,
                draw=none, fill=none, legend columns=5},
  bar width=3pt,
  ymax=40, ytick={0,5,10,15,20,25,30,35,40},
]
\addplot[fill=blue!70,   draw=blue!80!black]
  coordinates {(Comm.,15.0)(Dis.,7.5)(Emot.,20.0)(Exper.,17.5)(Expt.,10.0)(KnowC.,5.0)(SenRe.,25.0)};
\addplot[fill=teal!60,   draw=teal!80!black]
  coordinates {(Comm.,2.6)(Dis.,10.3)(Emot.,12.8)(Exper.,12.8)(Expt.,15.4)(KnowC.,17.9)(SenRe.,28.2)};
\addplot[fill=orange!70, draw=orange!80!black]
  coordinates {(Comm.,13.6)(Dis.,10.6)(Emot.,10.6)(Exper.,15.2)(Expt.,19.7)(KnowC.,4.5)(SenRe.,25.8)};
\addplot[fill=violet!60, draw=violet!80!black]
  coordinates {(Comm.,21.4)(Dis.,7.1)(Emot.,15.7)(Exper.,10.0)(Expt.,18.6)(KnowC.,5.7)(SenRe.,21.4)};
\addplot[fill=red!50,    draw=red!70!black]
  coordinates {(Comm.,4.3)(Dis.,6.4)(Emot.,8.5)(Exper.,8.5)(Expt.,36.2)(KnowC.,10.6)(SenRe.,25.5)};
\end{axis}
\end{tikzpicture}
    \caption{Code distribution per puzzle. \textit{Experimenting} peaks sharply for the Weird symbols puzzle (36\%); \textit{Knowledge Curation} is lowest for Nine-dots (4.5\%); \textit{Communicating} is highest for Three sons.}
\label{fig:codes:bypuzzle}
\end{subfigure}

\medskip

\begin{subfigure}[t]{0.48\linewidth}
\centering
\begin{tikzpicture}
\begin{axis}[
  codeaxis,
  ymax=30, ytick={0,5,10,15,20,25, 30},
  title={Professionals vs.\ students},
  legend entries={Professionals, Students},
]
\addplot[fill=blue!55, draw=blue!70!black]
  coordinates {(Comm.,12.0)(Dis.,12.8)(Emot.,13.7)(Exper.,15.4)(Expt.,15.4)(KnowC.,4.3)(SenRe.,26.5)};
\addplot[fill=red!45, draw=red!70!black, pattern=north east lines, pattern color=red!60]
  coordinates {(Comm.,13.1)(Dis.,4.8)(Emot.,13.1)(Exper.,10.3)(Expt.,24.1)(KnowC.,11.0)(SenRe.,23.4)};
\end{axis}
\end{tikzpicture}
\caption{Code distribution for professionals (written, all 5 puzzles, $n=117$) and students (written + think-aloud, $n=145$), each normalised to 100\%. Students score higher on \textit{Experimenting} and \textit{Knowledge Curation}; professionals score higher on \textit{Disposition} and \textit{Experience}.}
\label{fig:codes:groups}
\end{subfigure}
\hfill

\begin{subfigure}[t]{0.48\linewidth}
\centering
\begin{tikzpicture}
\begin{axis}[
  codeaxis,
  title={Think-aloud vs.\ written},
  legend entries={TA Three sons, TA Weird symbols, Written (avg.)},
  ymax=45, ytick={0,5,10,15,20,25,30,35,40,45},
]
\addplot[fill=violet!55, draw=violet!70!black]
  coordinates {(Comm.,26.1)(Dis.,4.3)(Emot.,13.0)(Exper.,6.5)(Expt.,23.9)(KnowC.,6.5)(SenRe.,19.6)};
\addplot[fill=teal!50, draw=teal!70!black, pattern=dots, pattern color=teal!70]
  coordinates {(Comm.,3.2)(Dis.,3.2)(Emot.,12.9)(Exper.,12.9)(Expt.,41.9)(KnowC.,16.1)(SenRe.,9.7)};
\addplot[fill=gray!40, draw=gray!60]
  coordinates {(Comm.,10.8)(Dis.,10.3)(Emot.,13.5)(Exper.,14.1)(Expt.,15.7)(KnowC.,7.0)(SenRe.,28.6)};
\end{axis}
\end{tikzpicture}
\caption{Think-aloud sessions compared to average written workbook codes. \textit{Communicating} is strongly elevated in the Three sons think-aloud; \textit{Experimenting} is elevated in the Weird symbols think-aloud, consistent with the written data.}
\label{fig:codes:methods}
\end{subfigure}

\caption{Code-document analysis of written workbook reflections and think-aloud sessions. Values are percentages of total coded quotations within each group. Category labels are abbreviated: \textit{Comm.}~=~\textit{Communicating}, \textit{Dis.}~=~\textit{Disposition}, \textit{Emot.}~=~\textit{Emotions}, \textit{Exper.}~=~\textit{Experience}, \textit{Expt.}~=~\textit{Experimenting}, \textit{KnowC.}~=~\textit{Knowledge Curation}, \textit{SenRe.}~=~\textit{Sensemaking / reflection-in-action}.}
\label{fig:codeanalysis}
\end{figure*}

The differences between professionals and students are most visible in the \textit{Disposition} category, as shown in Figure~\ref{fig:codes:groups}, where the expert codings are considerably higher: professionals reflect more on their self-confidence and disposition toward problem-solving, which may reflect greater meta-cognitive awareness of their own approach.
Students, by contrast, show higher \textit{Experimenting} and \textit{Knowledge Curation}, suggesting that novices engage in more overt trial-and-error and are more consciously aware of reaching for prior knowledge, while professionals may apply it more automatically.
For the Nine-dots puzzle, \textit{Knowledge Curation} is low across both groups (4.5\%), consistent with its character as a spatial insight task that requires abandoning rather than applying existing knowledge.
The Next-line puzzle shows the reverse, with the highest \textit{Knowledge Curation} scores in the data set, possibly showing participants using their prior experiences to curate new knowledge.

Puzzle-specific patterns in \textit{Experimenting} suggest that the degree of ambiguity in a puzzle is an important reason for exploratory behaviour (Figure~\ref{fig:codeanalysis}).

The Weird symbols puzzle shows the highest \textit{Experimenting} codes overall, while the Three sons puzzle produces the highest \textit{Communicating} codes, suggesting that puzzles in which participants come up with multiple valid solutions or engage in social reasoning more easily elicit discussion and explanation in the reflections.

The comparison between written reflections and think-aloud sessions, shown in Figure~\ref{fig:codes:methods}, shows differences in how participants articulate their puzzle-solving experience.
In the think-aloud sessions, \textit{Communicating} codes are noticeably higher, particularly for Three sons, where \textit{Communicating} is 26.1\%, suggesting that verbalising in real time makes it easier for participants to explain their reasoning step by step.
This is in contrast to the written reflections, where these codes tend to be more evenly distributed across categories.
\textit{Experimenting} remains prominent in the Weird symbols think-aloud (41.9\%), similar to the pattern seen in written expert reflections and indicating that this puzzle triggers exploratory behaviour.
What is perhaps most telling is what the think-alouds contain \textit{less} of: \textit{Disposition} and \textit{Knowledge Curation} codes are minimal, implying that when participants are actively engaged in solving, they are less inclined to step back and reflect on their own attitudes or consciously label the knowledge they are using.

\begin{takeaway}
Think-alouds and written reflections seem to capture related but distinct layers of the puzzle-solving experience: think-alouds surface immediate reasoning, while written reflections show more meta-cognitive reflection.
\end{takeaway}

Taken together, the coding results are broadly consistent with the Likert scores, and where they appear to diverge, the difference is methodological rather than contradictory.

\section{What this means for how we run the workshops in the future}
\label{puzzle:sec:discussion}

The closer look confirmed that the puzzles seem to be doing something real.
The participants experiment, they get stuck, they revise their thinking, and the debrief is where it comes together.
But looking more carefully also revealed things that we could not see from the front of the room.

\subsection{Reflection on what we did differently}

The workbooks gave us data we could not have collected from observation alone, but they also introduced friction.
The purpose of the printed workbooks was to be able to work completely offline to avoid distractions commonly associated with working behind a screen.
For solving the puzzles themselves, this approach works fine.
However, many participants did not answer all the questions in the reflection part, particularly the open ones.
We did not study the motivation behind non-response, but it is a well-known phenomenon that in paper-based surveys the non-response rate is higher.
This is mainly attributed to the higher cognitive load of answering open questions~\cite{Tourangeau2000, Krosnick2010}, the required effort to answer them~\cite{Dillman2014}, and the way surveys are designed~\cite{Groves2009}.
For most of the participants, English was not their native language, which may have added to this load.
It is worth considering whether an online format would reduce this problem.

\begin{takeaway}
A downside of using paper-based workbooks is high non-response to open questions.
\end{takeaway}

Providing the workbooks in a digital or hybrid format could possibly help mitigate this effect by 'forcing' participants to answer the open questions.

Think-aloud sessions seem to be the most revealing source of data.
Two pairs of students are not a large sample, however, what the think-aloud sessions gave us ---a moment-by-moment trace of how hypotheses form, get tested, and get abandoned--- is hard to obtain in any other way.

\subsection{Offering the workshop in a hybrid or digital form}

In order to mitigate the negative side effects of using paper based workbooks, we developed an open-source web application~\cite{doorn_2026_19426383}.
This application offers an environment in which participants can solve puzzles and researchers can collect data.

The use of a web application can have benefits in terms of ease of analysis.
However, solving the puzzles with pen and paper can lead to different solutions, for example with the Nine-dots puzzle, solutions on paper can lead to folding the paper, creating a cylinder shape, or Möbius strips.
This is not possible in a web application, it is therefore recommended to use a hybrid approach and provide participants with pen and paper.

\subsection{Limitations}

The study has no control group
or pre- and post-measurements, so causal claims about the effect of puzzles on competence are not possible.
The sample is small and non-random: seven students from one course, twelve professionals from one professional meetup.
The expert group is self-selected and likely more reflective than the average professional, which may have inflated observed differences; no demographic data were collected for them.
Think-aloud protocols introduce an observer effect, and verbalising reasoning aloud is itself a cognitive intervention.
During the Think-aloud sessions no debrief was given, making it harder to measure the learning effect between puzzles.
Self-reported scores are subject to social desirability bias and limited meta-cognitive access,
which can partly explain the low \textit{Emotions} ratings.
For most of the participants, English was not their native language, adding cognitive load to understanding the questions and phrasing the written reflections.

\section{Reflections and next steps}
\label{puzzle:sec:conclusion}

We set out to study whether puzzle-based activities can improve critical testing literacy and whether the behaviours we observed in workshops align with the P4TEST framework that motivated their design.

The data from the workbook sessions and think-aloud recordings indicate three things that we had observed anecdotally but could now be seen more clearly.
First, the participants self-describe that they are \textit{Experimenting} and using \textit{Sensemaking / reflection-in-action} during the puzzles, which are core components of the P4TEST framework
Second, professionals and students seem to engage with the same puzzles differently: professionals draw on prior experience and reflect on their own habits of mind with more ease, while students use more trial-and-error approaches and are more aware of reaching for knowledge they do not yet have.
Third, emotions are present and during puzzle solving, but participants seem to find them difficult to articulate in writing after the fact.
These three findings are consistent with Likert scores and qualitative coding, and they are consistent with what we observed in the eleven workshops.

This shows that the puzzles are not the intervention on their own.
What matters is the full sequence: framing the task, solving it, debriefing it, and connecting the experience to testing practice.
Of these, the debrief is the least designed; however, it is the most important part.
It is where learning happens and where the connection to software testing practice can be made.
Treating the debrief as a designed activity, rather than an open discussion, is the most significant change we can make to the workshop format.

The work ahead includes the following: designing a more structured debrief, using the developed web application to explore a hybrid approach of the workshops, and study whether the behaviours observed during puzzles transfer to software testing tasks.

\balance

{\RaggedRight\sloppy\addcontentsline{toc}{section}{References}\bibliographystyle{unsrtnat}\bibliography{bibliography}}

\begin{thebibliography}{19}
\providecommand{\natexlab}[1]{#1}
\providecommand{\url}[1]{\texttt{#1}}
\expandafter\ifx\csname urlstyle\endcsname\relax
  \providecommand{\doi}[1]{doi: #1}\else
  \providecommand{\doi}{doi: \begingroup \urlstyle{rm}\Url}\fi

\bibitem[Doorn et~al.(2023)Doorn, Vos, and Mar\'{\i}n]{DOORN2023102199}
Niels Doorn, Tanja E.~J. Vos, and Beatriz Mar\'{\i}n.
\newblock Towards understanding students' sensemaking of test case design.
\newblock \emph{Data \& Knowledge Engineering}, 146:\penalty0 102199, 2023.
\newblock ISSN 0169-023X.
\newblock \doi{https://doi.org/10.1016/j.datak.2023.102199}.
\newblock URL \url{https://www.sciencedirect.com/science/article/pii/S0169023X23000599}.

\bibitem[Doorn et~al.(2025)Doorn, Vos, Mar\'{\i}n, and van Diggelen]{doorn2025puzzle}
Niels Doorn, Tanja E.~J. Vos, Beatriz Mar\'{\i}n, and Migchiel van Diggelen.
\newblock Puzzle-based learning for developing software testing skills.
\newblock In \emph{Proceedings of the 2025 29th International Conference on Evaluation and Assessment in Software Engineering Companion}, EASE Companion '25, pages 202--212, New York, NY, USA, 2025. Association for Computing Machinery.
\newblock ISBN 9798400718328.
\newblock \doi{10.1145/3727967.3756829}.
\newblock URL \url{https://doi.org/10.1145/3727967.3756829}.

\bibitem[Vos et~al.(2026)Vos, Knaack, Mar\'{\i}n, Doorn, and van Vugt-Hag\`{e}]{vos2025P4TEST}
Tanja E.~J. Vos, Bart~Th. Knaack, Beatriz Mar\'{\i}n, Niels Doorn, and Nik\`{e} van Vugt-Hag\`{e}.
\newblock Teaching testing seriously in academia.
\newblock ENASE, 2026.

\bibitem[Falkner et~al.(2012)Falkner, Sooriamurthi, and Michalewicz]{FSM12}
Nickolas Falkner, Raja Sooriamurthi, and Zbigniew Michalewicz.
\newblock Teaching puzzle-based learning: Development of basic concepts.
\newblock \emph{Teaching Mathematics and Computer Science}, 10, 06 2012.
\newblock \doi{10.5485/TMCS.2012.0303}.

\bibitem[ste(1995)]{sternberg_nature_1995}
The nature of insight.
\newblock \emph{The nature of insight.}, pages xviii, 618--xviii, 618, 1995.
\newblock ISSN 0-262-19345-0 (Hardcover).

\bibitem[Doorn(2026)]{doorn_2026_19426383}
Niels Doorn.
\newblock Puzzle-based testing — web application, April 2026.
\newblock URL \url{https://doi.org/10.5281/zenodo.19426383}.

\bibitem[Loyd(1914)]{Loyd1914}
Sam Loyd.
\newblock \emph{Cyclopedia of 5000 Puzzles, Tricks, and Conundrums}.
\newblock The Lamb Publishing Company, 1914.
\newblock URL \url{https://archive.org/details/CyclopediaOfPuzzlesLoyd}.
\newblock Accessed: 2024-02-01.

\bibitem[Adams(2001)]{Adams2001}
James~L. Adams.
\newblock \emph{Conceptual Blockbusting: A Guide to Better Ideas}.
\newblock Perseus Publishing, Cambridge, MA, 4th edition, 2001.
\newblock ISBN 978-0738205373.

\bibitem[de~Bono(1970)]{deBono1970}
Edward de~Bono.
\newblock \emph{Lateral Thinking: Creativity Step by Step}.
\newblock Harper \& Row, New York, 1970.

\bibitem[Meyers and See(1990)]{MS90}
Leroy~F. Meyers and Richard See.
\newblock The census-taker problem.
\newblock \emph{Mathematics Magazine}, 63\penalty0 (2):\penalty0 86--88, 1990.
\newblock ISSN 0025570X, 19300980.
\newblock URL \url{http://www.jstor.org/stable/2691063}.

\bibitem[III et~al.(2014)III, Falkner, Sooriamurthi, and Michalewicz]{Meyeretal2014}
Edwin F.~Meyer III, Nickolas Falkner, Raja Sooriamurthi, and Zbigniew Michalewicz.
\newblock \emph{Guide to Teaching Puzzle-based Learning}.
\newblock Undergraduate Topics in Computer Science. Springer London, London, 1 edition, 2014.
\newblock ISBN 978-1-4471-6475-3.
\newblock \doi{10.1007/978-1-4471-6476-0}.

\bibitem[Kolb(1984)]{Kolb1984}
David~A Kolb.
\newblock \emph{Experiential Learning: Experience as the Source of Learning and Development}.
\newblock Prentice Hall, Englewood Cliffs, NJ, 1984.

\bibitem[Sweller(1976)]{Sweller1976}
John Sweller.
\newblock The effect of task complexity and sequence on rule learning and problem solving.
\newblock \emph{British Journal of Psychology}, 67\penalty0 (4):\penalty0 553--558, 1976.
\newblock \doi{10.1111/j.2044-8295.1976.tb01546.x}.

\bibitem[Li et~al.(2012)Li, Cohen, and Koedinger]{10.1007/978-3-642-30950-2_24}
Nan Li, William~W. Cohen, and Kenneth~R. Koedinger.
\newblock Problem order implications for learning transfer.
\newblock In Stefano~A. Cerri, William~J. Clancey, Giorgos Papadourakis, and Kitty Panourgia, editors, \emph{Intelligent Tutoring Systems}, pages 185--194, Berlin, Heidelberg, 2012. Springer Berlin Heidelberg.
\newblock ISBN 978-3-642-30950-2.

\bibitem[Doorn et~al.(2026)Doorn, Knaack, Vos, and Mar\'{\i}n]{doorn2026puzzledata}
Niels Doorn, Bart Knaack, Tanja E.~J. Vos, and Beatriz Mar\'{\i}n.
\newblock Research data: Puzzle-based learning of critical testing competences, 2026.
\newblock URL \url{https://doi.org/10.5281/zenodo.19278865}.

\bibitem[Tourangeau et~al.(2000)Tourangeau, Rips, and Rasinski]{Tourangeau2000}
Roger Tourangeau, Lance~J. Rips, and Kenneth Rasinski.
\newblock \emph{The Psychology of Survey Response}.
\newblock Cambridge University Press, Cambridge, 2000.

\bibitem[Krosnick and Presser(2010)]{Krosnick2010}
Jon~A. Krosnick and Stanley Presser.
\newblock Question and questionnaire design.
\newblock In Peter~V. Marsden and James~D. Wright, editors, \emph{Handbook of Survey Research}, pages 263--314. Emerald Group Publishing, 2 edition, 2010.

\bibitem[Dillman et~al.(2014)Dillman, Smyth, and Christian]{Dillman2014}
Don~A. Dillman, Jolene~D. Smyth, and Leah~Melani Christian.
\newblock \emph{Internet, Phone, Mail, and Mixed-Mode Surveys: The Tailored Design Method}.
\newblock Wiley, Hoboken, NJ, 4 edition, 2014.

\bibitem[Groves et~al.(2009)Groves, Fowler, Couper, Lepkowski, Singer, and Tourangeau]{Groves2009}
Robert~M. Groves, Floyd~J. Fowler, Mick~P. Couper, James~M. Lepkowski, Eleanor Singer, and Roger Tourangeau.
\newblock \emph{Survey Methodology}.
\newblock Wiley, Hoboken, NJ, 2 edition, 2009.

\end{thebibliography}

\end{document}